\documentclass[english,jmp, reprint, amsmath, amssymb]{revtex4-2}
\usepackage[T1]{fontenc}
\usepackage[latin9]{inputenc}
\usepackage{amssymb}
\usepackage{graphicx}
\usepackage{babel}
\begin{document}

\title{Josephson Diode Effect in Andreev Molecules}

\author{J.-D. Pillet}
\email{jean-damien.pillet@polytechnique.edu}
\author{S. Annabi, A. Peugeot, H. Riechert, E. Arrighi, J. Griesmar}
\author{L. Bretheau}
\email{landry.bretheau@polytechnique.edu}
\selectlanguage{english}

\affiliation{Laboratoire de Physique de la Mati\`ere condens\'ee, CNRS, Ecole Polytechnique, Institut Polytechnique de Paris, 91120 Palaiseau, France}

\begin{abstract}
We propose a new platform for observing the Josephson diode effect: the Andreev molecule. This nonlocal electronic state is hosted in circuits made of two closely spaced Josephson junctions, through the hybridization of the Andreev states. The Josephson diode effect occurs at the level of one individual junction while the other one generates the required time-reversal and spatial-inversion symmetry breaking. 
We present a microscopic description of this phenomenon based on fermionic Andreev states,  focusing on single channels in the short limit, and we compute both supercurrent and energy spectra. We demonstrate that the diode efficiency can be tuned by magnetic flux and the junctions transmissions, and can reach $45~\%$.
Going further, by analyzing the Andreev spectra, we demonstrate the key role played by the continuum, which consists of leaky Andreev states and is largely responsible for the critical current asymmetry. On top of proposing an 
experimentally accessible platform, this work elucidates the microscopic origin of the Josephson diode effect at the level of the fermionic Andreev states.

\end{abstract}

\maketitle

\section{Introduction}

In a Josephson diode (JD), the magnitude of the critical current depends on the direction in which the supercurrent flows. Although it looks exotic, this phenomenon is actually quite ubiquitous and appears in multiple superconducting circuits. For instance, it has been known for decades that asymmetric SQUIDs can behave as such, either when they are composed of  tunnel Josephson junctions and linear inductances \cite{fulton_quantum_1972,LeMasne2009} or when they are made of well-transmitted Josephson junctions \cite{DellaRocca2007,Bretheau2013a,Souto2022}.
The JD effect, and its close cousin the superconducting diode effect, have recently received renewed interest in the field of quantum materials, with experiments performed in superconducting films \cite{Ando2020,Bauriedl2022,Shin2021} and  Josephson junctions \cite{Bocquillon2017,Wu2022,Baumgartner2022,jeon_zero-field_2022,Pal2022,Baumgartner2022a,Trahms2023}. In both cases, the materials, which behave as a bulk superconductor or play the role of the junction's weak link, display internal symmetry breakings that originate from magnetic interactions (either intrinsic like spin-orbit or induced by a magnetic field)
or a non-centrosymmetric crystalline structure.
Examining the critical current asymmetry can actually serve as a practical and potent method for investigating the electronic characteristics of novel materials, such as detecting the presence and type (Rashba or Dresselhaus) of spin-orbit interaction~\cite{rasmussen_effects_2016}.

\begin{figure}
\includegraphics[width=1\columnwidth]{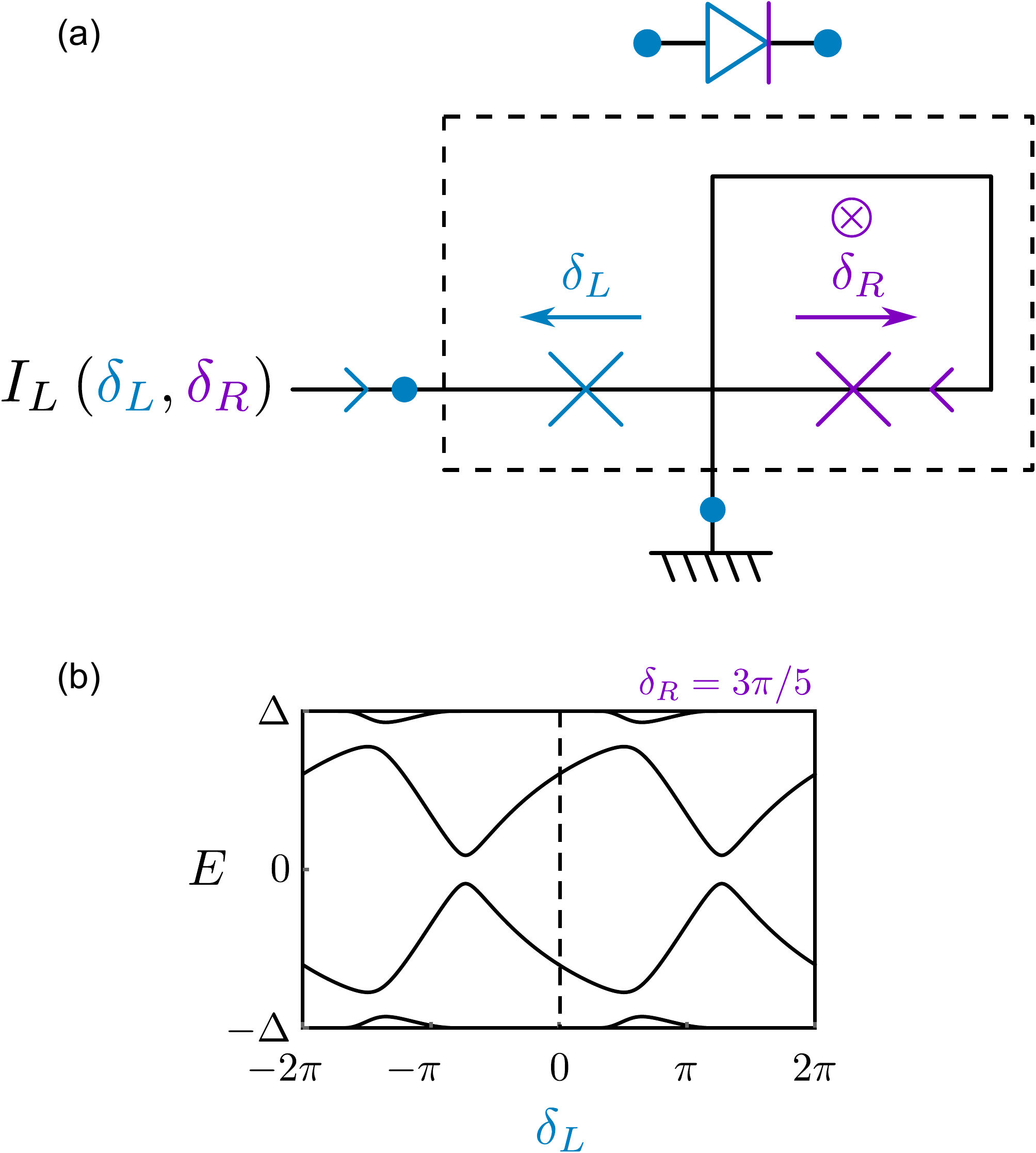}
\caption{\label{fig:Fig1} (a) Schematic of the proposed device for the observation of the Josephson diode effect in an Andreev molecule. The Josephson current of the left junction $I_L$ is measured as a function of the superconducting phase difference $\delta_R$ imposed across the right junction by a magnetic flux.
(b) Energy spectrum of an Andreev molecule, as a function of $\delta_L \in [-2\pi,2\pi]$, for $\delta_R=3\pi/5$. The spectral asymmetry with respect to $\delta_L=0$, which originates from time-reversal and inversion symmetry breakings, is responsible for the Josephson diode effect. Parameters used for calculations: 
$\tau_L=\tau_R=0.94$,
$l=\xi_0/2$ and $k_F=(2\times10^4+1/2)\pi/l$.}
\end{figure}

To observe the JD effect, it is required to break both spatial-inversion and time-reversal symmetries. In a symmetric system, the current-phase relation satisfies $I(-\delta)=-I(\delta)$, which indeed imposes $I_{c}^+=-I_{c}^-$ for the critical currents \textit{i.e.} no JD effect.
This broken-symmetry requirement is fulfilled in asymmetric SQUIDs at finite magnetic flux \cite{fulton_quantum_1972,LeMasne2009,DellaRocca2007,Bretheau2013a,Souto2022}, in the above-mentioned Josephson weak links~\cite{Bocquillon2017,Wu2022,Baumgartner2022,jeon_zero-field_2022,Pal2022,Baumgartner2022a,Trahms2023}, in multiterminal Josephson devices \cite{gupta_superconducting_2022,chiles_non-reciprocal_2022,zhang_reconfigurable_2023,melin_dc-josephson_2021} 
and in systems with finite Cooper pair momentum~\cite{davydova_universal_2022}.
In this work, we propose and investigate a new platform that can display the JD effect: the Andreev molecule.
It is composed of two closely-spaced well-transmitted Josephson junctions (JJ), which exhibit hybridization of their respective Andreev spectrum into a molecular state. Consequently, this system exhibits nonlocal Josephson effect, with the supercurrent flowing through each JJ depending on the superconducting phase difference across the other one~\cite{pillet_nonlocal_2019,pillet_scattering_2020,kornich_fine_2019,kornich_overlapping_2020,kocsis_strong_2023}. Another striking 
consequence of hybridization in Andreev molecules is the JD effect, which occurs at the level of one JJ due to the proximity of the other one. We present here a theoretical description of this phenomenon in terms of microscopic Andreev states, such an approach being sparsely explored in the literature on JD \cite{davydova_universal_2022}.
By computing the supercurrent as a function of the relevant parameters, we show that the JD effect on one JJ can be controlled by phase-biasing the other JJ and that its magnitude depends on the distance between the JJ as well as their respective transmission. The JD effect can thus be considered as a sensitive probe to characterize the degree of nonlocality of the Andreev molecule. 
Going further, we compute the energy spectra and investigate the respective role played by the Andreev bound states (ABS) and the continuum.
This microscopic analysis enables us to elucidate the mechanisms leading to symmetry breakings that manifest directly in the Andreev spectra and cause the JD effect.

\section{Modeling and symmetry analysis}
\label{sec_Mod}

Fig. \ref{fig:Fig1}a shows the schematics of the device, which implements a Josephson diode. It emphasizes that the device we consider is a two-terminal element,  fed on one side by a current $I_L$ and connected to the ground on the other side.
The circuit is primarily composed of a JJ based on a single-channel weak link of transmission $\tau_L$. This left JJ is in series with a second junction of transmission $\tau_R$, the right JJ, which is enclosed in a loop threaded by a magnetic flux.
In such a configuration, the supercurrents carried by the two junctions $I_L$ and $I_R$ can be different and the superconducting phase differences $\delta_L$ and $\delta_R$ can be controlled independently. In the following, we will study how the right JJ can influence the supercurrent $I_L$ through the left JJ and generate a critical current asymmetry.

We model this circuit using the Bogolubiov-de-Gennes formalism, where electrons obey the $2\times2$ Hamiltonian in the Nambu space
\begin{align}
H &= \left(\begin{array}{cc}
H_{0}+H_{\mathit{WL}} & \Delta(x) \\
\Delta^{*}(x) & -H_{0}-H_{\mathit{WL}}
\end{array}\right).\label{eq:Hamiltonian}
\end{align}
Here $H_{0}=\frac{-\hbar^{2}}{2m}\partial_{x}^{2}-\mu$ is the single particle energy, with $m$ the electron mass and $\mu$ the chemical potential. 
The scattering at the two weak links, located at $x=\pm l/2$, is modeled by $H_{\mathit{WL}}=U_{L}\delta\left(x+l/2\right)+U_{R}\delta\left(x-l/2\right)$, with amplitudes $U_{L/R}$ related to the transmissions $\tau_{L/R}=1/[1+(U_{L/R}/\hbar v_F)^2]$, where $v_F$ is the Fermi velocity. For simplicity we indeed use Dirac $\delta$-functions for the scatterers, which are appropriate for weak links of length shorter than the superconducting coherence length $\xi_0$.
Electron pairing in each superconductor is described by the step function
\begin{align}
\Delta(x) &= \begin{cases}
\Delta e^{i\delta_{L}} & \text{if }x<-l/2\\
\Delta & \text{if }|x|<l/2\\
\Delta e^{i\delta_{R}} & \text{if }x>l/2
\end{cases}\label{eq:offdiag}
\end{align}
where the superconducting gap amplitude $\Delta$ is considered constant along the whole device. 
In the following, we will use this model to compute the eigenspectrum of $H$, as a function of the relevant parameters $\delta_L$, $\delta_R$, $\tau_L$, $\tau_R$ and $l$. One typically finds a spectrum composed of discrete ABS at energies $E_{ABS} \in ]-\Delta,\Delta[$ and a continuum of scattering states at energies $|E|\ge \Delta$.
More details about the calculations can be found in reference~\cite{pillet_nonlocal_2019}.

Within this model, the device behavior  depends crucially on the relative distance $l/\xi_0$ between the junctions. When $l \gg \xi_0$, each junction hosts independent ABS and does not exhibit nonlocal Josephson effect. On the contrary, when $l\sim\xi_0$ the ABS of the left and right junction couple and hybridize. This leads to avoided crossings in the Andreev spectrum when two ABS become degenerate.
As the coupling between ABS increases, the size of these avoided crossings grows, potentially pushing some ABS outside of the superconducting gap. This can result in a spectrum where the total number of ABS varies with the phase, as illustrated in Fig. \ref{fig:Fig1}b where one varies $\delta_L \in [-2\pi,2\pi]$ with a constant $\delta_R =3\pi/5$. This calculation, and all of the followings, is performed for a fixed distance $l=\xi_0/2$, where the hybridization between ABS is strong~\cite{pillet_nonlocal_2019} and where we expect a large JD effect.

Strikingly, the ABS spectrum shown in Fig.~\ref{fig:Fig1}b is asymmetric with respect to $\delta_L=0$ or $\pi$. This  originates from both spatial-inversion and time-reversal symmetry breaking at the left JJ level. The Hamiltonian $H(\delta_L,\delta_R)$ of Eq.~\eqref{eq:Hamiltonian} indeed satisfies the following symmetry constraints. 
First, it trivially breaks local space-inversion symmetry 
$\mathcal{I} H(\delta_L,\delta_R)\mathcal{I}^{-1} \ne H(-\delta_L,\delta_R)$, where 
$\mathcal{I}$
is the unitary inversion operator. 
Second, it has a global time-reversal symmetry 
$\mathcal{T} H(\delta_L,\delta_R)\mathcal{T}^{-1} = H(-\delta_L,-\delta_R)$, the antiunitary operator $\mathcal{T}$ being complex conjugation. This causes $I_{L/R}(-\delta_L,-\delta_R)=-I_{L/R}(\delta_L,\delta_R)$. 
But crucially, time-reversal symmetry can locally be broken at the level of the left JJ. Considering  $\delta_R \ne 0~(\mathrm{mod}~\pi)$ as a fixed parameter, one indeed gets $\mathcal{T} H(\delta_L,\delta_R)\mathcal{T}^{-1} \ne H(-\delta_L,\delta_R)$. This results in $I_L(-\delta_L) \neq -I_L(\delta_L)$ and one can get  $I_{Lc}^+\neq -I_{Lc}^-$. From the perspective of the left JJ only, the required symmetries are broken, which allows it to behave as a JD.
Interestingly, these symmetry breakings are achieved in a nonlocal way due to the spatial extension of the ground state wavefunction over the two JJ. Though the phase $\delta_R$ is set across the right JJ, it affects the current flowing through the left one, which causes the JD effect.

\begin{figure*}
\includegraphics[width=0.8\textwidth]{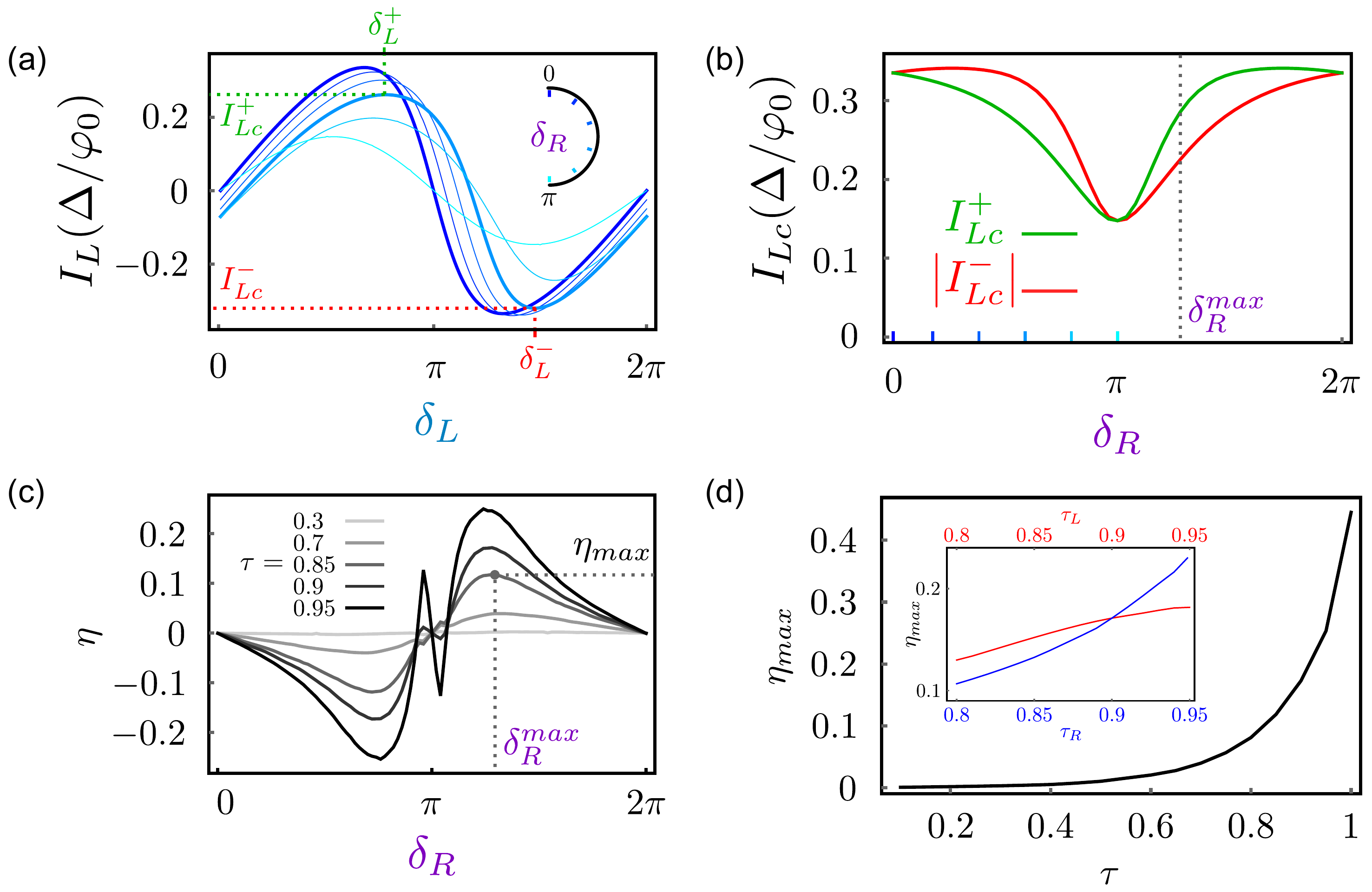}
\caption{\label{fig:Fig2} (a) Current-phase relation $I_L(\delta_L)$ of the left junction, for phases $\delta_R=n\pi/5$ ($n \in [\![0,5]\!]$) and for symmetric transmissions $\tau=\tau_{L/R}=0.85$. 
The extreme values $I_{Lc}^+$ and $I_{Lc}^-$ and the corresponding phases $\delta_{L}^+$ and $\delta_{L}^-$ are indicated by dotted lines in the case $\delta_R=3\pi/5$, where the CPR is clearly asymmetric. 
(b) Critical currents of the left junction $I_{Lc}^\pm$ as a function of $\delta_R$, still for $\tau=0.85$. The JD effect is finite when $\delta_R \ne 0~(\mathrm{mod}~\pi)$.
The ticks indicate at which phases $\delta_R$ the curves of Fig \ref{fig:Fig2}a have been calculated. (c) Josephson diode efficiency $\eta$ as a function of $\delta_R$, for different transmissions $\tau$. In (c) and (d), the $\delta_R^{max}$ mark corresponds to the maximum efficiency. (d) Maximum efficiency $\eta_{max}$ as a function of transmission $\tau$. Inset: $\eta_{max}$ as a function of $\tau_R$ (resp. $\tau_L$), for fixed $\tau_{L(R)} = 0.9$. We have here introduced an asymmetry between the left and right junction $\tau_L\neq\tau_R$. Parameters used for calculations: 
$l=\xi_0/2$ and $k_F=(2\times10^4+1/2)\pi/l$.}
\end{figure*}

\section{Supercurrent and diode effect}
\label{sec_supercurrent}

To study the JD effect, we now derive the supercurrent flowing through the left junction at zero temperature
\begin{align}
I_L = \frac{1}{\varphi_0}\sum_{E_{ABS}<0}\frac{\partial E_{ABS}}{\partial \delta_L} + I_L^{cont.}
\label{eq_current}
\end{align}
where the first term corresponds to the contribution of the negative-energy ABS ($\varphi_0$ is the reduced flux quantum). The second term is the current carried by the continuum, which is obtained by integrating the contribution of all negative-energy scattering states~\cite{pillet_nonlocal_2019}.
Fig. \ref{fig:Fig2}a shows such a current $I_L$, as a function of $\delta_L$. This local current-phase relation (CPR) is plotted for different values of $\delta_R$, for symmetric transmissions $\tau=\tau_{L/R}=0.85$. The CPR of the left JJ depends on $\delta_R$, thus demonstrating the nonlocal Josephson effect~\cite{pillet_nonlocal_2019,pillet_scattering_2020,kornich_fine_2019,kornich_overlapping_2020}.
More importantly here, the minimum and maximum values of the CPR, also known as the negative and positive critical current (red and green dashed lines), can be different in magnitude. Indeed, when $0<\delta_R<\pi$, one gets $I_{Lc}^+ \ne -I_{Lc}^-$, i.e. a finite JD effect. 
Moreover, the critical currents are achieved at asymmetric phases $\delta_{L}^+ \ne -\delta_{L}^-~(\mathrm{mod}~2\pi)$.
The CPR thus shows a shift in phase with a finite supercurrent at $\delta_L=0$.
Both phenomena---JD effect and $\varphi_0$-junction physics---originate from symmetry breaking. In contrast, when time-reversal symmetry is restored at $\delta_R = 0~(\mathrm{mod}~\pi)$, one finds $I_L(\delta_L=0)=0$ and $I_{Lc}^+=-I_{Lc}^-$ (darkest and lightest blue curves).

The dependence of the critical currents $I_{Lc}^+$ and $-I_{Lc}^-$ as a function of $\delta_R$ is shown in Fig.~\ref{fig:Fig2}b for $\tau=0.85$.
The critical current of the left JJ modulates with the phase of the right JJ, which is a direct signature of nonlocal Josephson effect, as recently observed in Andreev molecules based on InAs/Al heterostructures~\cite{matsuo_observation_2022,haxell2023demonstration}.
On top of that, Fig. \ref{fig:Fig2}b directly exhibits critical current asymmetry. The JD effect is indeed finite for all phase values but $\delta_R = 0~(\mathrm{mod}~\pi)$, where the system is time-reversal invariant. Interestingly, one finds the relation $I_{Lc}^+ (\delta_R)=-I_{Lc}^- (-\delta_R)$, a direct consequence of the global time-reversal symmetry mentioned in Section~\ref{sec_Mod}.

To further characterize the JD effect, we introduce the dimensionless diode efficiency
\begin{align}
\eta=\frac{I_{Lc}^+-|I_{Lc}^-|}{I_{Lc}^++|I_{Lc}^-|},
\end{align}
such that $\eta=0$ means no JD effect and $\eta=1$ corresponds to an ideal Josephson diode. The diode efficiency is shown in Fig.~\ref{fig:Fig2}c as a function of $\delta_R$, for different transmissions $\tau$, in the case of symmetric junctions. The efficiency is an odd function of $\delta_R$ due to global time-reversal symmetry. It modulates periodically and cancels at $\delta_R=0$ and $\pi$. It reaches its maximal value $\eta_{max}$ at an intermediate phase that depends on the transmission $\tau$. The maximum efficiency grows exponentially with $\tau$, as shown in Fig.~\ref{fig:Fig2}d. It remains below $\sim 2 ~\%$ for $\tau<0.5$, while it reaches $\sim 45~\%$ for $\tau=1$. Finally, we explored the influence of transmission asymmetry $\tau_L \ne \tau_R$, as illustrated by the inset of Fig. \ref{fig:Fig2}d. We find that the JD effect is favored by higher transmissions for both junctions.

We have thus shown that the JD effect occurs 
in Andreev molecules, with large efficiencies that correspond to critical current asymmetries as big as $I_{Lc}^+/|I_{Lc}^-| \simeq 265~\%$. Let's now try to elucidate its microscopic origin, at the level of the fermionic Andreev states.

\begin{figure*}
\includegraphics[width=1\textwidth]{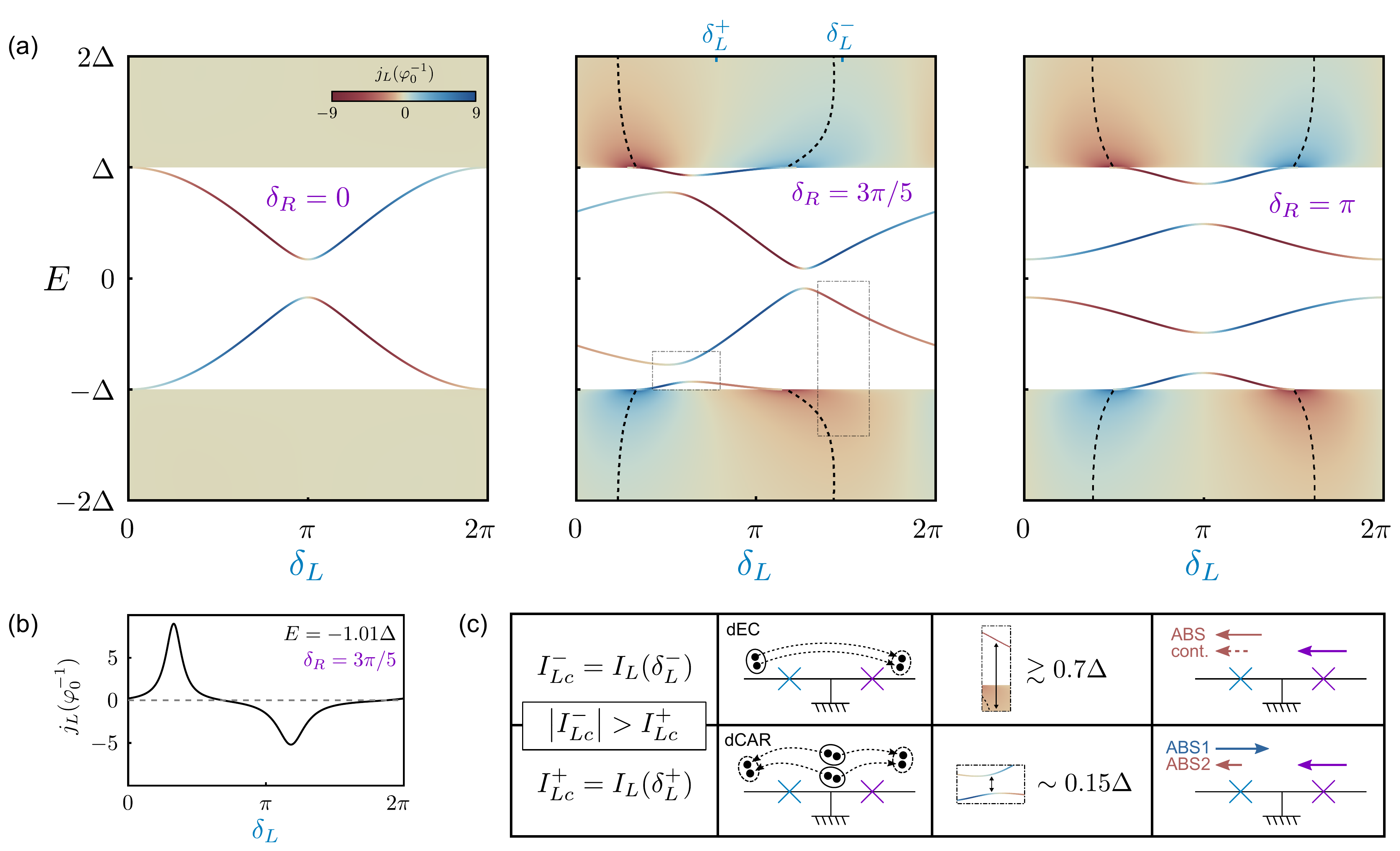}
\caption{\label{fig:Fig3} 
(a) Combined Andreev spectra for $\delta_R=0$, $3\pi/5$ and $\pi$. For $|E| < \Delta$: energy of the ABS, $E_{ABS}(\delta_L)$, where the line color corresponds to the magnitude of the ABS supercurrent (see Eq.~\ref{eq_current}), 
encoded from red (negative) to blue (positive) in arbitrary units.
For $|E| \ge \Delta$: supercurrent spectral density $j_L(E,\delta_L)$ through the left JJ. The local extrema are highlighted as dotted black lines and correspond to the leaky Andreev states. 
$\delta_L^+$ and $\delta_L^-$ are the phases at which the supercurrent $I_L$ reaches its extreme values $I_{Lc}^+$ and $I_{Lc}^-$. 
(b) Supercurrent spectral density $j_L(\delta_L)$ for $\delta_R=3\pi/5$ and $E=-1.01\Delta$. The broad resonances correspond to leaky Andreev states that carry positive and negative supercurrent. 
(c) Table comparing the microscopic situation at  $\delta_L^-$ (top) and $\delta_L^+$ (bottom).
In the first case, dEC is the main mechanism carrying the supercurrent while dCAR dominates in the second one. The imbalance of these two mechanisms leads to avoided crossings of different magnitude around $\delta_L=\pm\delta_R$ (boxes from the central Andreev spectrum in (a)), which is responsible for the overall asymmetry of the spectrum. The schematic on the right illustrates the contribution of each microscopic state (ABS or continuum) to the critical current $I_{Lc}^\pm$. The purple arrow indicates the direction of $I_R$ for $\delta_R=3\pi/5$. 
Parameters used for calculations: 
same as Fig. \ref{fig:Fig1}.
}
\end{figure*}

\section{Microscopic analysis}

For our microscopic analysis, we compute both the density of states and the supercurrent spectral density $j_L$ of the left JJ, in the energy domain. In Fig.~\ref{fig:Fig3}a, we introduce a new compact representation that allows to visualize both quantities at the same time. For energies $|E| \ge \Delta$, we plot the color-coded supercurrent density $j_L$ of the continuum. For energies $|E|<\Delta$, we plot the ABS energies while the supercurrent they carry is encoded in color. Such combined Andreev spectra are shown in Fig.~\ref{fig:Fig3}a, as a function of energy $E$ and local phase $\delta_L$, for three different phases $\delta_R$ across the right JJ. We have chosen here large symmetric transmissions $\tau_L=\tau_R=0.94$, for which we have demonstrated a strong JD effect.
At $\delta_R=0$ (left panel), the two JJ do not couple and we recover the well-known ABS spectrum of a single channel in the short limit. In that case, the continuum does not contribute to the supercurrent \cite{PhysRevLett.66.3056}, \emph{i.e.} $j_L=0$. 

The spectra are dramatically changed when $\delta_R=3\pi/5$ or $\pi$ (central and right panels of Fig.~\ref{fig:Fig3}a). In such cases, the ABS of the left and right junctions hybridize, resulting in the emergence of avoided crossings around the degeneracy points (at $\delta_L=\pm\delta_R$ for $\tau_L=\tau_R$). Due to this level-repulsion, ABS can be expelled into the continuum and transform into Andreev scattering states. These "leaky" Andreev states are no longer discrete but acquire a finite width inside the continuum. They appear as broad resonances in the current density $j_L$ (see linecut in Fig. \ref{fig:Fig3}b), with a width that increases as $|E|$ departs from the gap edge at $|E|=\Delta$ until they fade away for $|E| \to \infty$. 
Similarly to the ABS energy, these resonances, which are highlighted as dashed lines in the Andreev spectra, depend on $\delta_L$, thereby carrying a supercurrent. Strikingly, the leaky Andreev states appear precisely at phases $\delta_L$ where the ABS vanish in the continuum. Using this compact representation, we could thus make explicit the connection between ABS and leaky Andreev states. This explains why the continuum can carry significant supercurrent in Andreev molecules.

Going further, we now study the Andreev spectra symmetry that determines the appearance of the JD effect. At $\delta_R =0$ or $\pi$ (left and right panels of Fig.~\ref{fig:Fig3}a), the overall symmetry of the spectra with respect to $\delta_L$ indicates the absence of the JD effect. This is consistent with both our global symmetry analysis performed in Section~\ref{sec_Mod} and our results from Section~\ref{sec_supercurrent}. Interestingly, at $\delta_R =\pi$, the continuum carries a significant supercurrent, opposite to the one carried by the ABS. This results in a largely reduced critical current, as can be seen in Fig.~\ref{fig:Fig2}b. However, this does not lead to a critical current asymmetry since time-reversal symmetry is preserved. When $\delta_R \ne 0~(\mathrm{mod}~\pi)$, the Andreev spectra are no longer symmetrical with respect to $\delta_L$, as exemplified by the central panel of Fig.~\ref{fig:Fig3}a computed at $\delta_R=3\pi/5$. In that case, the positive and  negative critical currents are such that $|I_{Lc}^-|>I_{Lc}^+$. These currents are reached at phases $\delta_L^+$ and $\delta_L^-$, where the Andreev spectrum is completely different, with two negative energy ABS at $\delta_L^+$ and a single one at $\delta_L^-$. 

This asymmetry takes root in the different microscopic mechanisms responsible for the ABS hybridization (see Fig.~\ref{fig:Fig3}c). 
At $\delta_L^+$, the dominant mechanism is the double crossed Andreev reflection (dCAR), while at $\delta_L^-$ it involves double elastic cotunneling (dEC) of Cooper pairs \cite{freyn_production_2011,feinberg_quartets_2015,pillet_scattering_2020}. Indeed, at $\delta_L=\delta_L^\pm$, which is in the vicinity of $\pm\delta_R$, currents $I_L$ and $I_R$ are counter-propagating (resp. co-propagating) via dCAR (resp. dEC). These two mechanisms are not equally probable, hence the asymmetry, especially at large transmission. When $\tau\sim 1$, dEC is very likely while the dCAR probability vanishes to zero due to spatial translational symmetry that results in conserved momenta~\cite{pillet_scattering_2020}. The avoided crossing is therefore much larger in the vicinity of $\delta_L^-$, where dEC is the dominant mechanism. It is so strong that one pair of ABS is expelled into the continuum and morphs into leaky Andreev states. Consequently, the positive and negative critical currents are carried by very different fermionic states. Indeed at $\delta_L=\delta_L^-$, both the ABS and the continuum contribute to $I_{Lc}^-$, in the same direction (red arrows corresponding to negative current in Fig.~\ref{fig:Fig3}c, last column). In contrast at $\delta_L=\delta_L^+$, the continuum contribution is negligible and the current is carried by two different ABS, with opposite direction (blue and red arrows), resulting in a largely reduced critical current $I_{Lc}^+$. Interestingly, it seems that this JD effect can be related to the current flow in the right JJ (purple arrow in Fig.~\ref{fig:Fig3}c, last column). We see that the critical current is smaller when $I_L$ and $I_R$ counter-propagate, as if the right JJ was inducing a friction-like effect, impeding the supercurrent flow through the left JJ. Finally, we have seen that the emergence of the JD effect is linked to a key contribution of the continuum to the supercurrent, though we are in the short junction limit. Such a phenomenon
seems intimately related to symmetry breakings, as
recently discussed in Ref.~\cite{davydova_universal_2022} in the context of finite Cooper pair momentum. Remarkably, we find this same notion in the Andreev molecule.

\section{Concluding remarks}

In conclusion, we have demonstrated the occurrence of the Josephson diode effect in Andreev molecules.
This quantum phenomenon is a ground state property and results from the nonlocality of the Andreev molecule wavefunction.
Our findings reveal that the critical current of a junction can depend on the phase across a neighboring junction
as well as the relative current flow,
unequivocally establishing the presence of the JD effect.
The efficiency $\eta$ of this effect increases with the transmission of weak links, and can reach $45~\%$. Although disorder 
is likely to diminish $\eta$ significantly, such a large effect should be readily accessible experimentally.
By performing both a symmetry and a microscopic study, we could investigate the origin of the JD effect. In particular, by analyzing the Andreev spectra, we could demonstrate the key role played by the continuum composed of leaky Andreev states and their connection to the ABS.
We hope that our work will motivate further exploration of the JD effect, both theoretically and experimentally, employing a similar microscopic approach by examining the spectrum of Andreev states and the role of the continuum. Such investigations could potentially lead to an improved understanding of the underlying mechanisms responsible for the JD effect in exotic quantum materials.
On the other hand, it would be interesting to investigate this physics in the context of superconducting quantum dots, where Coulomb interaction plays a crucial role. Moreover, we anticipate that intricate architectures comprising chains of several junctions, i.e. Andreev polymers, could display both long-range nonlocal Josephson effect and enhance Josephson diode effect. Finally, the asymmetric Josephson potential engineered in Andreev molecules could be harnessed to implement non-reciprocal quantum devices operating at microwave frequencies such as circulators and quantum limited amplifiers~\cite{sliwa_reconfigurable_2015,frattini_3-wave_2017,frattini_optimizing_2018}.

During the preparation of the manuscript, we became aware of
a related experimental work 
\cite{matsuo2023josephson}.

\subsection*{Acknowledgements}

JDP acknowledges support of Agence Nationale de la Recherche through grant ANR-20-CE47-0003.
LB acknowledges support of the European Research Council (ERC) under the European Union's Horizon 2020 research and innovation programme (grant agreement No. 947707).


\begin{thebibliography}{35}%
\makeatletter
\providecommand \@ifxundefined [1]{%
 \@ifx{#1\undefined}
}%
\providecommand \@ifnum [1]{%
 \ifnum #1\expandafter \@firstoftwo
 \else \expandafter \@secondoftwo
 \fi
}%
\providecommand \@ifx [1]{%
 \ifx #1\expandafter \@firstoftwo
 \else \expandafter \@secondoftwo
 \fi
}%
\providecommand \natexlab [1]{#1}%
\providecommand \enquote  [1]{``#1''}%
\providecommand \bibnamefont  [1]{#1}%
\providecommand \bibfnamefont [1]{#1}%
\providecommand \citenamefont [1]{#1}%
\providecommand \href@noop [0]{\@secondoftwo}%
\providecommand \href [0]{\begingroup \@sanitize@url \@href}%
\providecommand \@href[1]{\@@startlink{#1}\@@href}%
\providecommand \@@href[1]{\endgroup#1\@@endlink}%
\providecommand \@sanitize@url [0]{\catcode `\\12\catcode `\$12\catcode
  `\&12\catcode `\#12\catcode `\^12\catcode `\_12\catcode `\%12\relax}%
\providecommand \@@startlink[1]{}%
\providecommand \@@endlink[0]{}%
\providecommand \url  [0]{\begingroup\@sanitize@url \@url }%
\providecommand \@url [1]{\endgroup\@href {#1}{\urlprefix }}%
\providecommand \urlprefix  [0]{URL }%
\providecommand \Eprint [0]{\href }%
\providecommand \doibase [0]{https://doi.org/}%
\providecommand \selectlanguage [0]{\@gobble}%
\providecommand \bibinfo  [0]{\@secondoftwo}%
\providecommand \bibfield  [0]{\@secondoftwo}%
\providecommand \translation [1]{[#1]}%
\providecommand \BibitemOpen [0]{}%
\providecommand \bibitemStop [0]{}%
\providecommand \bibitemNoStop [0]{.\EOS\space}%
\providecommand \EOS [0]{\spacefactor3000\relax}%
\providecommand \BibitemShut  [1]{\csname bibitem#1\endcsname}%
\let\auto@bib@innerbib\@empty
\bibitem [{\citenamefont {Fulton}\ \emph {et~al.}(1972)\citenamefont {Fulton},
  \citenamefont {Dunkleberger},\ and\ \citenamefont
  {Dynes}}]{fulton_quantum_1972}%
  \BibitemOpen
  \bibfield  {author} {\bibinfo {author} {\bibfnamefont {T.~A.}\ \bibnamefont
  {Fulton}}, \bibinfo {author} {\bibfnamefont {L.~N.}\ \bibnamefont
  {Dunkleberger}},\ and\ \bibinfo {author} {\bibfnamefont {R.~C.}\ \bibnamefont
  {Dynes}},\ }\href {https://doi.org/10.1103/PhysRevB.6.855} {\bibfield
  {journal} {\bibinfo  {journal} {Physical Review B}\ }\textbf {\bibinfo
  {volume} {6}},\ \bibinfo {pages} {855} (\bibinfo {year} {1972})} \BibitemShut {NoStop}%
\bibitem [{\citenamefont {{Le Masne}}(2009)}]{LeMasne2009}%
  \BibitemOpen
  \bibfield  {author} {\bibinfo {author} {\bibfnamefont {Q.}~\bibnamefont {{Le
  Masne}}},\ }\emph {\bibinfo {title} {{Asymmetric current fluctuations and
  Andreev states probed with a Josephson junction}}},\ \href@noop {} {Ph.D.
  thesis} (\bibinfo {year} {2009})\BibitemShut {NoStop}%
\bibitem [{\citenamefont {{Della Rocca}}\ \emph {et~al.}(2007)\citenamefont
  {{Della Rocca}}, \citenamefont {Chauvin}, \citenamefont {Huard},
  \citenamefont {Pothier}, \citenamefont {Esteve},\ and\ \citenamefont
  {Urbina}}]{DellaRocca2007}%
  \BibitemOpen
  \bibfield  {author} {\bibinfo {author} {\bibfnamefont {M.~L.}\ \bibnamefont
  {{Della Rocca}}}, \bibinfo {author} {\bibfnamefont {M.}~\bibnamefont
  {Chauvin}}, \bibinfo {author} {\bibfnamefont {B.}~\bibnamefont {Huard}},
  \bibinfo {author} {\bibfnamefont {H.}~\bibnamefont {Pothier}}, \bibinfo
  {author} {\bibfnamefont {D.}~\bibnamefont {Esteve}},\ and\ \bibinfo {author}
  {\bibfnamefont {C.}~\bibnamefont {Urbina}},\ }\href
  {https://doi.org/10.1103/PhysRevLett.99.127005} {\bibfield  {journal}
  {\bibinfo  {journal} {Phys. Rev. Lett.}\ }\textbf {\bibinfo {volume} {99}},\
  \bibinfo {pages} {127005} (\bibinfo {year} {2007})}\BibitemShut {NoStop}%
\bibitem [{\citenamefont {Bretheau}(2013)}]{Bretheau2013a}%
  \BibitemOpen
  \bibfield  {author} {\bibinfo {author} {\bibfnamefont {L.}~\bibnamefont
  {Bretheau}},\ }\emph {\bibinfo {title} {{Localized Excitations in
  Superconducting Atomic Contacts : Probing the Andreev Doublet}}},\ \href
  {http://hal.archives-ouvertes.fr/tel-00772851/} {\bibinfo {type} {Phd
  thesis}} (\bibinfo {year} {2013})\BibitemShut {NoStop}%
\bibitem [{\citenamefont {Souto}\ \emph {et~al.}(2022)\citenamefont {Souto},
  \citenamefont {Leijnse},\ and\ \citenamefont {Schrade}}]{Souto2022}%
  \BibitemOpen
  \bibfield  {author} {\bibinfo {author} {\bibfnamefont {R.~S.}\ \bibnamefont
  {Souto}}, \bibinfo {author} {\bibfnamefont {M.}~\bibnamefont {Leijnse}},\
  and\ \bibinfo {author} {\bibfnamefont {C.}~\bibnamefont {Schrade}},\ }\href
  {https://doi.org/10.1103/PhysRevLett.129.267702} {\bibfield  {journal}
  {\bibinfo  {journal} {Phys. Rev. Lett.}\ }\textbf {\bibinfo {volume} {129}},\
  \bibinfo {pages} {1} (\bibinfo {year} {2022})} \BibitemShut {NoStop}%
\bibitem [{\citenamefont {Ando}\ \emph {et~al.}(2020)\citenamefont {Ando},
  \citenamefont {Miyasaka}, \citenamefont {Li}, \citenamefont {Ishizuka},
  \citenamefont {Arakawa}, \citenamefont {Shiota}, \citenamefont {Moriyama},
  \citenamefont {Yanase},\ and\ \citenamefont {Ono}}]{Ando2020}%
  \BibitemOpen
  \bibfield  {author} {\bibinfo {author} {\bibfnamefont {F.}~\bibnamefont
  {Ando}}, \bibinfo {author} {\bibfnamefont {Y.}~\bibnamefont {Miyasaka}},
  \bibinfo {author} {\bibfnamefont {T.}~\bibnamefont {Li}}, \bibinfo {author}
  {\bibfnamefont {J.}~\bibnamefont {Ishizuka}}, \bibinfo {author}
  {\bibfnamefont {T.}~\bibnamefont {Arakawa}}, \bibinfo {author} {\bibfnamefont
  {Y.}~\bibnamefont {Shiota}}, \bibinfo {author} {\bibfnamefont
  {T.}~\bibnamefont {Moriyama}}, \bibinfo {author} {\bibfnamefont
  {Y.}~\bibnamefont {Yanase}},\ and\ \bibinfo {author} {\bibfnamefont
  {T.}~\bibnamefont {Ono}},\ }\href {https://doi.org/10.1038/s41586-020-2590-4}
  {\bibfield  {journal} {\bibinfo  {journal} {Nature}\ }\textbf {\bibinfo
  {volume} {584}},\ \bibinfo {pages} {373} (\bibinfo {year}
  {2020})}\BibitemShut {NoStop}%
\bibitem [{\citenamefont {Bauriedl}\ \emph {et~al.}(2022)\citenamefont
  {Bauriedl}, \citenamefont {B{\"{a}}uml}, \citenamefont {Fuchs}, \citenamefont
  {Baumgartner}, \citenamefont {Paulik}, \citenamefont {Bauer}, \citenamefont
  {Lin}, \citenamefont {Lupton}, \citenamefont {Taniguchi}, \citenamefont
  {Watanabe}, \citenamefont {Strunk},\ and\ \citenamefont
  {Paradiso}}]{Bauriedl2022}%
  \BibitemOpen
  \bibfield  {author} {\bibinfo {author} {\bibfnamefont {L.}~\bibnamefont
  {Bauriedl}}, \bibinfo {author} {\bibfnamefont {C.}~\bibnamefont
  {B{\"{a}}uml}}, \bibinfo {author} {\bibfnamefont {L.}~\bibnamefont {Fuchs}},
  \bibinfo {author} {\bibfnamefont {C.}~\bibnamefont {Baumgartner}}, \bibinfo
  {author} {\bibfnamefont {N.}~\bibnamefont {Paulik}}, \bibinfo {author}
  {\bibfnamefont {J.~M.}\ \bibnamefont {Bauer}}, \bibinfo {author}
  {\bibfnamefont {K.~Q.}\ \bibnamefont {Lin}}, \bibinfo {author} {\bibfnamefont
  {J.~M.}\ \bibnamefont {Lupton}}, \bibinfo {author} {\bibfnamefont
  {T.}~\bibnamefont {Taniguchi}}, \bibinfo {author} {\bibfnamefont
  {K.}~\bibnamefont {Watanabe}}, \bibinfo {author} {\bibfnamefont
  {C.}~\bibnamefont {Strunk}},\ and\ \bibinfo {author} {\bibfnamefont
  {N.}~\bibnamefont {Paradiso}},\ }\href
  {https://doi.org/10.1038/s41467-022-31954-5} {\bibfield  {journal} {\bibinfo
  {journal} {Nat. Commun.}\ }\textbf {\bibinfo {volume} {13}},\ \bibinfo
  {pages} {1} (\bibinfo {year} {2022})}, \BibitemShut {NoStop}%
\bibitem [{\citenamefont {Shin}\ \emph {et~al.}(2021)\citenamefont {Shin},
  \citenamefont {Son}, \citenamefont {Yun}, \citenamefont {Park}, \citenamefont
  {Zhang}, \citenamefont {Shin}, \citenamefont {Park},\ and\ \citenamefont
  {Kim}}]{Shin2021}%
  \BibitemOpen
  \bibfield  {author} {\bibinfo {author} {\bibfnamefont {J.}~\bibnamefont
  {Shin}}, \bibinfo {author} {\bibfnamefont {S.}~\bibnamefont {Son}}, \bibinfo
  {author} {\bibfnamefont {J.}~\bibnamefont {Yun}}, \bibinfo {author}
  {\bibfnamefont {G.}~\bibnamefont {Park}}, \bibinfo {author} {\bibfnamefont
  {K.}~\bibnamefont {Zhang}}, \bibinfo {author} {\bibfnamefont {Y.~J.}\
  \bibnamefont {Shin}}, \bibinfo {author} {\bibfnamefont {J.-G.}\ \bibnamefont
  {Park}},\ and\ \bibinfo {author} {\bibfnamefont {D.}~\bibnamefont {Kim}},\
  }\href {http://arxiv.org/abs/2111.05627} (\bibinfo {year}
  {2021}),\ \Eprint {https://arxiv.org/abs/2111.05627} {arXiv:2111.05627
  }\BibitemShut {NoStop}%
\bibitem [{\citenamefont {Bocquillon}\ \emph {et~al.}(2017)\citenamefont
  {Bocquillon}, \citenamefont {Deacon}, \citenamefont {Wiedenmann},
  \citenamefont {Leubner}, \citenamefont {Klapwijk}, \citenamefont
  {Br{\"{u}}ne}, \citenamefont {Ishibashi}, \citenamefont {Buhmann},\ and\
  \citenamefont {Molenkamp}}]{Bocquillon2017}%
  \BibitemOpen
  \bibfield  {author} {\bibinfo {author} {\bibfnamefont {E.}~\bibnamefont
  {Bocquillon}}, \bibinfo {author} {\bibfnamefont {R.~S.}\ \bibnamefont
  {Deacon}}, \bibinfo {author} {\bibfnamefont {J.}~\bibnamefont {Wiedenmann}},
  \bibinfo {author} {\bibfnamefont {P.}~\bibnamefont {Leubner}}, \bibinfo
  {author} {\bibfnamefont {T.~M.}\ \bibnamefont {Klapwijk}}, \bibinfo {author}
  {\bibfnamefont {C.}~\bibnamefont {Br{\"{u}}ne}}, \bibinfo {author}
  {\bibfnamefont {K.}~\bibnamefont {Ishibashi}}, \bibinfo {author}
  {\bibfnamefont {H.}~\bibnamefont {Buhmann}},\ and\ \bibinfo {author}
  {\bibfnamefont {L.~W.}\ \bibnamefont {Molenkamp}},\ }\href
  {https://doi.org/10.1038/nnano.2016.159} {\bibfield  {journal} {\bibinfo
  {journal} {Nat. Nanotechnol.}\ }\textbf {\bibinfo {volume} {12}},\ \bibinfo
  {pages} {137} (\bibinfo {year} {2017})} \BibitemShut {NoStop}%
\bibitem [{\citenamefont {Wu}\ \emph {et~al.}(2022)\citenamefont {Wu},
  \citenamefont {Wang}, \citenamefont {Xu}, \citenamefont {Sivakumar},
  \citenamefont {Pasco}, \citenamefont {Filippozzi}, \citenamefont {Parkin},
  \citenamefont {Zeng}, \citenamefont {McQueen},\ and\ \citenamefont
  {Ali}}]{Wu2022}%
  \BibitemOpen
  \bibfield  {author} {\bibinfo {author} {\bibfnamefont {H.}~\bibnamefont
  {Wu}}, \bibinfo {author} {\bibfnamefont {Y.}~\bibnamefont {Wang}}, \bibinfo
  {author} {\bibfnamefont {Y.}~\bibnamefont {Xu}}, \bibinfo {author}
  {\bibfnamefont {P.~K.}\ \bibnamefont {Sivakumar}}, \bibinfo {author}
  {\bibfnamefont {C.}~\bibnamefont {Pasco}}, \bibinfo {author} {\bibfnamefont
  {U.}~\bibnamefont {Filippozzi}}, \bibinfo {author} {\bibfnamefont {S.~S.}\
  \bibnamefont {Parkin}}, \bibinfo {author} {\bibfnamefont {Y.~J.}\
  \bibnamefont {Zeng}}, \bibinfo {author} {\bibfnamefont {T.}~\bibnamefont
  {McQueen}},\ and\ \bibinfo {author} {\bibfnamefont {M.~N.}\ \bibnamefont
  {Ali}},\ }\href {https://doi.org/10.1038/s41586-022-04504-8} {\bibfield
  {journal} {\bibinfo  {journal} {Nature}\ }\textbf {\bibinfo {volume} {604}},\
  \bibinfo {pages} {653} (\bibinfo {year} {2022})}\BibitemShut {NoStop}%
\bibitem [{\citenamefont {Baumgartner}\ \emph
  {et~al.}(2022{\natexlab{a}})\citenamefont {Baumgartner}, \citenamefont
  {Fuchs}, \citenamefont {Costa}, \citenamefont {Reinhardt}, \citenamefont
  {Gronin}, \citenamefont {Gardner}, \citenamefont {Lindemann}, \citenamefont
  {Manfra}, \citenamefont {{Faria Junior}}, \citenamefont {Kochan},
  \citenamefont {Fabian}, \citenamefont {Paradiso},\ and\ \citenamefont
  {Strunk}}]{Baumgartner2022}%
  \BibitemOpen
  \bibfield  {author} {\bibinfo {author} {\bibfnamefont {C.}~\bibnamefont
  {Baumgartner}}, \bibinfo {author} {\bibfnamefont {L.}~\bibnamefont {Fuchs}},
  \bibinfo {author} {\bibfnamefont {A.}~\bibnamefont {Costa}}, \bibinfo
  {author} {\bibfnamefont {S.}~\bibnamefont {Reinhardt}}, \bibinfo {author}
  {\bibfnamefont {S.}~\bibnamefont {Gronin}}, \bibinfo {author} {\bibfnamefont
  {G.~C.}\ \bibnamefont {Gardner}}, \bibinfo {author} {\bibfnamefont
  {T.}~\bibnamefont {Lindemann}}, \bibinfo {author} {\bibfnamefont {M.~J.}\
  \bibnamefont {Manfra}}, \bibinfo {author} {\bibfnamefont {P.~E.}\
  \bibnamefont {{Faria Junior}}}, \bibinfo {author} {\bibfnamefont
  {D.}~\bibnamefont {Kochan}}, \bibinfo {author} {\bibfnamefont
  {J.}~\bibnamefont {Fabian}}, \bibinfo {author} {\bibfnamefont
  {N.}~\bibnamefont {Paradiso}},\ and\ \bibinfo {author} {\bibfnamefont
  {C.}~\bibnamefont {Strunk}},\ }\href
  {https://doi.org/10.1038/s41565-021-01009-9} {\bibfield  {journal} {\bibinfo
  {journal} {Nat. Nanotechnol.}\ }\textbf {\bibinfo {volume} {17}},\ \bibinfo
  {pages} {39} (\bibinfo {year} {2022}{\natexlab{a}})}\BibitemShut {NoStop}%
\bibitem [{\citenamefont {Jeon}\ \emph {et~al.}(2022)\citenamefont {Jeon},
  \citenamefont {Kim}, \citenamefont {Yoon}, \citenamefont {Jeon},
  \citenamefont {Han}, \citenamefont {Cottet}, \citenamefont {Kontos},\ and\
  \citenamefont {Parkin}}]{jeon_zero-field_2022}%
  \BibitemOpen
  \bibfield  {author} {\bibinfo {author} {\bibfnamefont {K.-R.}\ \bibnamefont
  {Jeon}}, \bibinfo {author} {\bibfnamefont {J.-K.}\ \bibnamefont {Kim}},
  \bibinfo {author} {\bibfnamefont {J.}~\bibnamefont {Yoon}}, \bibinfo {author}
  {\bibfnamefont {J.-C.}\ \bibnamefont {Jeon}}, \bibinfo {author}
  {\bibfnamefont {H.}~\bibnamefont {Han}}, \bibinfo {author} {\bibfnamefont
  {A.}~\bibnamefont {Cottet}}, \bibinfo {author} {\bibfnamefont
  {T.}~\bibnamefont {Kontos}},\ and\ \bibinfo {author} {\bibfnamefont
  {S.~S.~P.}\ \bibnamefont {Parkin}},\ }\href
  {https://doi.org/10.1038/s41563-022-01300-7} {\bibfield  {journal} {\bibinfo
  {journal} {Nature Materials}\ }\textbf {\bibinfo {volume} {21}},\ \bibinfo
  {pages} {1008} (\bibinfo {year} {2022})} \BibitemShut {NoStop}%
\bibitem [{\citenamefont {Pal}\ \emph {et~al.}(2022)\citenamefont {Pal},
  \citenamefont {Chakraborty}, \citenamefont {Sivakumar}, \citenamefont
  {Davydova}, \citenamefont {Gopi}, \citenamefont {Pandeya}, \citenamefont
  {Krieger}, \citenamefont {Zhang}, \citenamefont {Date}, \citenamefont {Ju},
  \citenamefont {Yuan}, \citenamefont {Schr{\"{o}}ter}, \citenamefont {Fu},\
  and\ \citenamefont {Parkin}}]{Pal2022}%
  \BibitemOpen
  \bibfield  {author} {\bibinfo {author} {\bibfnamefont {B.}~\bibnamefont
  {Pal}}, \bibinfo {author} {\bibfnamefont {A.}~\bibnamefont {Chakraborty}},
  \bibinfo {author} {\bibfnamefont {P.~K.}\ \bibnamefont {Sivakumar}}, \bibinfo
  {author} {\bibfnamefont {M.}~\bibnamefont {Davydova}}, \bibinfo {author}
  {\bibfnamefont {A.~K.}\ \bibnamefont {Gopi}}, \bibinfo {author}
  {\bibfnamefont {A.~K.}\ \bibnamefont {Pandeya}}, \bibinfo {author}
  {\bibfnamefont {J.~A.}\ \bibnamefont {Krieger}}, \bibinfo {author}
  {\bibfnamefont {Y.}~\bibnamefont {Zhang}}, \bibinfo {author} {\bibfnamefont
  {M.}~\bibnamefont {Date}}, \bibinfo {author} {\bibfnamefont {S.}~\bibnamefont
  {Ju}}, \bibinfo {author} {\bibfnamefont {N.}~\bibnamefont {Yuan}}, \bibinfo
  {author} {\bibfnamefont {N.~B.}\ \bibnamefont {Schr{\"{o}}ter}}, \bibinfo
  {author} {\bibfnamefont {L.}~\bibnamefont {Fu}},\ and\ \bibinfo {author}
  {\bibfnamefont {S.~S.}\ \bibnamefont {Parkin}},\ }\href
  {https://doi.org/10.1038/s41567-022-01699-5} {\bibfield  {journal} {\bibinfo
  {journal} {Nat. Phys.}\ }\textbf {\bibinfo {volume} {18}},\ \bibinfo {pages}
  {1228} (\bibinfo {year} {2022})} \BibitemShut {NoStop}%
\bibitem [{\citenamefont {Baumgartner}\ \emph
  {et~al.}(2022{\natexlab{b}})\citenamefont {Baumgartner}, \citenamefont
  {Fuchs}, \citenamefont {Costa}, \citenamefont {Pic{\'{o}}-Cort{\'{e}}s},
  \citenamefont {Reinhardt}, \citenamefont {Gronin}, \citenamefont {Gardner},
  \citenamefont {Lindemann}, \citenamefont {Manfra}, \citenamefont {{Faria
  Junior}}, \citenamefont {Kochan}, \citenamefont {Fabian}, \citenamefont
  {Paradiso},\ and\ \citenamefont {Strunk}}]{Baumgartner2022a}%
  \BibitemOpen
  \bibfield  {author} {\bibinfo {author} {\bibfnamefont {C.}~\bibnamefont
  {Baumgartner}}, \bibinfo {author} {\bibfnamefont {L.}~\bibnamefont {Fuchs}},
  \bibinfo {author} {\bibfnamefont {A.}~\bibnamefont {Costa}}, \bibinfo
  {author} {\bibfnamefont {J.}~\bibnamefont {Pic{\'{o}}-Cort{\'{e}}s}},
  \bibinfo {author} {\bibfnamefont {S.}~\bibnamefont {Reinhardt}}, \bibinfo
  {author} {\bibfnamefont {S.}~\bibnamefont {Gronin}}, \bibinfo {author}
  {\bibfnamefont {G.~C.}\ \bibnamefont {Gardner}}, \bibinfo {author}
  {\bibfnamefont {T.}~\bibnamefont {Lindemann}}, \bibinfo {author}
  {\bibfnamefont {M.~J.}\ \bibnamefont {Manfra}}, \bibinfo {author}
  {\bibfnamefont {P.~E.}\ \bibnamefont {{Faria Junior}}}, \bibinfo {author}
  {\bibfnamefont {D.}~\bibnamefont {Kochan}}, \bibinfo {author} {\bibfnamefont
  {J.}~\bibnamefont {Fabian}}, \bibinfo {author} {\bibfnamefont
  {N.}~\bibnamefont {Paradiso}},\ and\ \bibinfo {author} {\bibfnamefont
  {C.}~\bibnamefont {Strunk}},\ }\bibfield  {journal} {\bibinfo  {journal} {J.
  Phys. Condens. Matter}\ }\textbf {\bibinfo {volume} {34}}
  (\bibinfo {year} {2022}{\natexlab{b}}) \BibitemShut {NoStop}%
\bibitem [{\citenamefont {Trahms}\ \emph {et~al.}(2023)\citenamefont {Trahms},
  \citenamefont {Melischek}, \citenamefont {Steiner}, \citenamefont {Mahendru},
  \citenamefont {Tamir}, \citenamefont {Bogdanoff}, \citenamefont {Peters},
  \citenamefont {Reecht}, \citenamefont {Winkelmann}, \citenamefont {von
  Oppen},\ and\ \citenamefont {Franke}}]{Trahms2023}%
  \BibitemOpen
  \bibfield  {author} {\bibinfo {author} {\bibfnamefont {M.}~\bibnamefont
  {Trahms}}, \bibinfo {author} {\bibfnamefont {L.}~\bibnamefont {Melischek}},
  \bibinfo {author} {\bibfnamefont {J.~F.}\ \bibnamefont {Steiner}}, \bibinfo
  {author} {\bibfnamefont {B.}~\bibnamefont {Mahendru}}, \bibinfo {author}
  {\bibfnamefont {I.}~\bibnamefont {Tamir}}, \bibinfo {author} {\bibfnamefont
  {N.}~\bibnamefont {Bogdanoff}}, \bibinfo {author} {\bibfnamefont
  {O.}~\bibnamefont {Peters}}, \bibinfo {author} {\bibfnamefont
  {G.}~\bibnamefont {Reecht}}, \bibinfo {author} {\bibfnamefont {C.~B.}\
  \bibnamefont {Winkelmann}}, \bibinfo {author} {\bibfnamefont
  {F.}~\bibnamefont {von Oppen}},\ and\ \bibinfo {author} {\bibfnamefont
  {K.~J.}\ \bibnamefont {Franke}},\ }\bibfield  {journal} {\bibinfo  {journal}
  {Nature}\ }\textbf {\bibinfo {volume} {615}},
  (\bibinfo {year} {2023}) \BibitemShut {NoStop}%
\bibitem [{\citenamefont {Rasmussen}\ \emph {et~al.}(2016)\citenamefont
  {Rasmussen}, \citenamefont {Danon}, \citenamefont {Suominen}, \citenamefont
  {Nichele}, \citenamefont {Kjaergaard},\ and\ \citenamefont
  {Flensberg}}]{rasmussen_effects_2016}%
  \BibitemOpen
  \bibfield  {author} {\bibinfo {author} {\bibfnamefont {A.}~\bibnamefont
  {Rasmussen}}, \bibinfo {author} {\bibfnamefont {J.}~\bibnamefont {Danon}},
  \bibinfo {author} {\bibfnamefont {H.}~\bibnamefont {Suominen}}, \bibinfo
  {author} {\bibfnamefont {F.}~\bibnamefont {Nichele}}, \bibinfo {author}
  {\bibfnamefont {M.}~\bibnamefont {Kjaergaard}},\ and\ \bibinfo {author}
  {\bibfnamefont {K.}~\bibnamefont {Flensberg}},\ }\href
  {https://doi.org/10.1103/PhysRevB.93.155406} {\bibfield  {journal} {\bibinfo
  {journal} {Physical Review B}\ }\textbf {\bibinfo {volume} {93}},\ \bibinfo
  {pages} {155406} (\bibinfo {year} {2016})} \BibitemShut {NoStop}%
\bibitem [{\citenamefont {Gupta}\ \emph {et~al.}(2022)\citenamefont {Gupta},
  \citenamefont {Graziano}, \citenamefont {Pendharkar}, \citenamefont {Dong},
  \citenamefont {Dempsey}, \citenamefont {Palmstr{\o}m},\ and\ \citenamefont
  {Pribiag}}]{gupta_superconducting_2022}%
  \BibitemOpen
  \bibfield  {author} {\bibinfo {author} {\bibfnamefont {M.}~\bibnamefont
  {Gupta}}, \bibinfo {author} {\bibfnamefont {G.~V.}\ \bibnamefont {Graziano}},
  \bibinfo {author} {\bibfnamefont {M.}~\bibnamefont {Pendharkar}}, \bibinfo
  {author} {\bibfnamefont {J.~T.}\ \bibnamefont {Dong}}, \bibinfo {author}
  {\bibfnamefont {C.~P.}\ \bibnamefont {Dempsey}}, \bibinfo {author}
  {\bibfnamefont {C.}~\bibnamefont {Palmstr{\o}m}},\ and\ \bibinfo {author}
  {\bibfnamefont {V.~S.}\ \bibnamefont {Pribiag}},\ }\href
  {https://doi.org/10.48550/arXiv.2206.08471}(\bibinfo {year}
  {2022}),\ \bibinfo {note} {arXiv:2206.08471}\BibitemShut {NoStop}%
\bibitem [{\citenamefont {Chiles}\ \emph {et~al.}(2022)\citenamefont {Chiles},
  \citenamefont {Arnault}, \citenamefont {Chen}, \citenamefont {Larson},
  \citenamefont {Zhao}, \citenamefont {Watanabe}, \citenamefont {Taniguchi},
  \citenamefont {Amet},\ and\ \citenamefont
  {Finkelstein}}]{chiles_non-reciprocal_2022}%
  \BibitemOpen
  \bibfield  {author} {\bibinfo {author} {\bibfnamefont {J.}~\bibnamefont
  {Chiles}}, \bibinfo {author} {\bibfnamefont {E.~G.}\ \bibnamefont {Arnault}},
  \bibinfo {author} {\bibfnamefont {C.-C.}\ \bibnamefont {Chen}}, \bibinfo
  {author} {\bibfnamefont {T.~F.~Q.}\ \bibnamefont {Larson}}, \bibinfo {author}
  {\bibfnamefont {L.}~\bibnamefont {Zhao}}, \bibinfo {author} {\bibfnamefont
  {K.}~\bibnamefont {Watanabe}}, \bibinfo {author} {\bibfnamefont
  {T.}~\bibnamefont {Taniguchi}}, \bibinfo {author} {\bibfnamefont
  {F.}~\bibnamefont {Amet}},\ and\ \bibinfo {author} {\bibfnamefont
  {G.}~\bibnamefont {Finkelstein}},\ }\href
  {https://doi.org/10.48550/arXiv.2210.02644}(\bibinfo {year}
  {2022}), \ \bibinfo {note} {arXiv:2210.02644
 }\BibitemShut {NoStop}%
\bibitem [{\citenamefont {Zhang}\ \emph {et~al.}(2023)\citenamefont {Zhang},
  \citenamefont {Ahari}, \citenamefont {Rashid}, \citenamefont {de~Coster},
  \citenamefont {Taniguchi}, \citenamefont {Watanabe}, \citenamefont {Gilbert},
  \citenamefont {Samarth},\ and\ \citenamefont
  {Kayyalha}}]{zhang_reconfigurable_2023}%
  \BibitemOpen
  \bibfield  {author} {\bibinfo {author} {\bibfnamefont {F.}~\bibnamefont
  {Zhang}}, \bibinfo {author} {\bibfnamefont {M.~T.}\ \bibnamefont {Ahari}},
  \bibinfo {author} {\bibfnamefont {A.~S.}\ \bibnamefont {Rashid}}, \bibinfo
  {author} {\bibfnamefont {G.~J.}\ \bibnamefont {de~Coster}}, \bibinfo {author}
  {\bibfnamefont {T.}~\bibnamefont {Taniguchi}}, \bibinfo {author}
  {\bibfnamefont {K.}~\bibnamefont {Watanabe}}, \bibinfo {author}
  {\bibfnamefont {M.~J.}\ \bibnamefont {Gilbert}}, \bibinfo {author}
  {\bibfnamefont {N.}~\bibnamefont {Samarth}},\ and\ \bibinfo {author}
  {\bibfnamefont {M.}~\bibnamefont {Kayyalha}},\ }\href
  {https://doi.org/10.48550/arXiv.2301.05081}(\bibinfo {year}
  {2023}), \ \bibinfo {note}
  {arXiv:2301.05081}\BibitemShut {NoStop}%
\bibitem [{\citenamefont {M{\'e}lin}(2021)}]{melin_dc-josephson_2021}%
  \BibitemOpen
  \bibfield  {author} {\bibinfo {author} {\bibfnamefont {R.}~\bibnamefont
  {M{\'e}lin}},\ }\href {https://doi.org/10.48550/arXiv.2103.03519}  (\bibinfo {year} {2021}),\ \bibinfo {note} {arXiv:2103.03519}\BibitemShut {NoStop}%
\bibitem [{\citenamefont {Davydova}\ \emph {et~al.}(2022)\citenamefont
  {Davydova}, \citenamefont {Prembabu},\ and\ \citenamefont
  {Fu}}]{davydova_universal_2022}%
  \BibitemOpen
  \bibfield  {author} {\bibinfo {author} {\bibfnamefont {M.}~\bibnamefont
  {Davydova}}, \bibinfo {author} {\bibfnamefont {S.}~\bibnamefont {Prembabu}},\
  and\ \bibinfo {author} {\bibfnamefont {L.}~\bibnamefont {Fu}},\ }\href
  {https://doi.org/10.1126/sciadv.abo0309} {\bibfield  {journal} {\bibinfo
  {journal} {Science Advances}\ }\textbf {\bibinfo {volume} {8}}, (\bibinfo {year} {2022})} \BibitemShut {NoStop}%
\bibitem [{\citenamefont {Pillet}\ \emph {et~al.}(2019)\citenamefont {Pillet},
  \citenamefont {Benzoni}, \citenamefont {Griesmar}, \citenamefont {Smirr},\
  and\ \citenamefont {Girit}}]{pillet_nonlocal_2019}%
  \BibitemOpen
  \bibfield  {author} {\bibinfo {author} {\bibfnamefont {J.-D.}\ \bibnamefont
  {Pillet}}, \bibinfo {author} {\bibfnamefont {V.}~\bibnamefont {Benzoni}},
  \bibinfo {author} {\bibfnamefont {J.}~\bibnamefont {Griesmar}}, \bibinfo
  {author} {\bibfnamefont {J.-L.}\ \bibnamefont {Smirr}},\ and\ \bibinfo
  {author} {\bibfnamefont {{\c C}.~{\"O}.}\ \bibnamefont {Girit}},\ }\href
  {https://doi.org/10.1021/acs.nanolett.9b02686} {\bibfield  {journal}
  {\bibinfo  {journal} {Nano Letters}\ }\textbf {\bibinfo {volume} {19}},\
  \bibinfo {pages} {7138} (\bibinfo {year} {2019})} \BibitemShut {NoStop}%
\bibitem [{\citenamefont {Pillet}\ \emph {et~al.}(2020)\citenamefont {Pillet},
  \citenamefont {Benzoni}, \citenamefont {Griesmar}, \citenamefont {Smirr},\
  and\ \citenamefont {Girit}}]{pillet_scattering_2020}%
  \BibitemOpen
  \bibfield  {author} {\bibinfo {author} {\bibfnamefont {J.-D.}\ \bibnamefont
  {Pillet}}, \bibinfo {author} {\bibfnamefont {V.}~\bibnamefont {Benzoni}},
  \bibinfo {author} {\bibfnamefont {J.}~\bibnamefont {Griesmar}}, \bibinfo
  {author} {\bibfnamefont {J.-L.}\ \bibnamefont {Smirr}},\ and\ \bibinfo
  {author} {\bibfnamefont {{\c C}.}~\bibnamefont {Girit}},\ }\href
  {https://doi.org/10.21468/SciPostPhysCore.2.2.009} {\bibfield  {journal}
  {\bibinfo  {journal} {SciPost Physics Core}\ }\textbf {\bibinfo {volume}
  {2}},\ \bibinfo {pages} {009} (\bibinfo {year} {2020})}\BibitemShut {NoStop}%
\bibitem [{\citenamefont {Kornich}\ \emph {et~al.}(2019)\citenamefont
  {Kornich}, \citenamefont {Barakov},\ and\ \citenamefont
  {Nazarov}}]{kornich_fine_2019}%
  \BibitemOpen
  \bibfield  {author} {\bibinfo {author} {\bibfnamefont {V.}~\bibnamefont
  {Kornich}}, \bibinfo {author} {\bibfnamefont {H.~S.}\ \bibnamefont
  {Barakov}},\ and\ \bibinfo {author} {\bibfnamefont {Y.~V.}\ \bibnamefont
  {Nazarov}},\ }\href {https://doi.org/10.1103/PhysRevResearch.1.033004}
  {\bibfield  {journal} {\bibinfo  {journal} {Physical Review Research}\
  }\textbf {\bibinfo {volume} {1}},\ \bibinfo {pages} {033004} (\bibinfo {year}
  {2019})}\BibitemShut
  {NoStop}%
\bibitem [{\citenamefont {Kornich}\ \emph {et~al.}(2020)\citenamefont
  {Kornich}, \citenamefont {Barakov},\ and\ \citenamefont
  {Nazarov}}]{kornich_overlapping_2020}%
  \BibitemOpen
  \bibfield  {author} {\bibinfo {author} {\bibfnamefont {V.}~\bibnamefont
  {Kornich}}, \bibinfo {author} {\bibfnamefont {H.~S.}\ \bibnamefont
  {Barakov}},\ and\ \bibinfo {author} {\bibfnamefont {Y.~V.}\ \bibnamefont
  {Nazarov}},\ }\href@noop {} {\bibfield  {journal} {\bibinfo  {journal}
  {Physical Review B}\ }\textbf {\bibinfo {volume} {101}},\ \bibinfo {pages}
  {195430} (\bibinfo {year} {2020})}\BibitemShut {NoStop}%
\bibitem [{\citenamefont {Kocsis}\ \emph {et~al.}(2023)\citenamefont {Kocsis},
  \citenamefont {Scher{\"u}bl}, \citenamefont {F{\"u}l{\"o}p}, \citenamefont
  {Makk},\ and\ \citenamefont {Csonka}}]{kocsis_strong_2023}%
  \BibitemOpen
  \bibfield  {author} {\bibinfo {author} {\bibfnamefont {M.}~\bibnamefont
  {Kocsis}}, \bibinfo {author} {\bibfnamefont {Z.}~\bibnamefont
  {Scher{\"u}bl}}, \bibinfo {author} {\bibfnamefont {G.}~\bibnamefont
  {F{\"u}l{\"o}p}}, \bibinfo {author} {\bibfnamefont {P.}~\bibnamefont
  {Makk}},\ and\ \bibinfo {author} {\bibfnamefont {S.}~\bibnamefont {Csonka}},\
  }\href {https://doi.org/10.48550/arXiv.2303.14842}  (\bibinfo {year} {2023}),\ \bibinfo {note}
  {arXiv:2303.14842}\BibitemShut {NoStop}%
\bibitem [{\citenamefont {Matsuo}\ \emph {et~al.}(2022)\citenamefont {Matsuo},
  \citenamefont {Lee}, \citenamefont {Chang}, \citenamefont {Sato},
  \citenamefont {Ueda}, \citenamefont {Palmstr{\o}m},\ and\ \citenamefont
  {Tarucha}}]{matsuo_observation_2022}%
  \BibitemOpen
  \bibfield  {author} {\bibinfo {author} {\bibfnamefont {S.}~\bibnamefont
  {Matsuo}}, \bibinfo {author} {\bibfnamefont {J.~S.}\ \bibnamefont {Lee}},
  \bibinfo {author} {\bibfnamefont {C.-Y.}\ \bibnamefont {Chang}}, \bibinfo
  {author} {\bibfnamefont {Y.}~\bibnamefont {Sato}}, \bibinfo {author}
  {\bibfnamefont {K.}~\bibnamefont {Ueda}}, \bibinfo {author} {\bibfnamefont
  {C.~J.}\ \bibnamefont {Palmstr{\o}m}},\ and\ \bibinfo {author} {\bibfnamefont
  {S.}~\bibnamefont {Tarucha}},\ }\href
  {https://doi.org/10.1038/s42005-022-00994-0} {\bibfield  {journal} {\bibinfo
  {journal} {Communications Physics}\ }\textbf {\bibinfo {volume} {5}},\
  \bibinfo {pages} {1} (\bibinfo {year} {2022})}\BibitemShut {NoStop}%
\bibitem [{\citenamefont {Haxell}\ \emph {et~al.}(2023)\citenamefont {Haxell},
  \citenamefont {Coraiola}, \citenamefont {Hinderling}, \citenamefont {ten
  Kate}, \citenamefont {Sabonis}, \citenamefont {Svetogorov}, \citenamefont
  {Belzig}, \citenamefont {Cheah}, \citenamefont {Krizek}, \citenamefont
  {Schott}, \citenamefont {Wegscheider},\ and\ \citenamefont
  {Nichele}}]{haxell2023demonstration}%
  \BibitemOpen
  \bibfield  {author} {\bibinfo {author} {\bibfnamefont {D.~Z.}\ \bibnamefont
  {Haxell}}, \bibinfo {author} {\bibfnamefont {M.}~\bibnamefont {Coraiola}},
  \bibinfo {author} {\bibfnamefont {M.}~\bibnamefont {Hinderling}}, \bibinfo
  {author} {\bibfnamefont {S.~C.}\ \bibnamefont {ten Kate}}, \bibinfo {author}
  {\bibfnamefont {D.}~\bibnamefont {Sabonis}}, \bibinfo {author} {\bibfnamefont
  {A.~E.}\ \bibnamefont {Svetogorov}}, \bibinfo {author} {\bibfnamefont
  {W.}~\bibnamefont {Belzig}}, \bibinfo {author} {\bibfnamefont
  {E.}~\bibnamefont {Cheah}}, \bibinfo {author} {\bibfnamefont
  {F.}~\bibnamefont {Krizek}}, \bibinfo {author} {\bibfnamefont
  {R.}~\bibnamefont {Schott}}, \bibinfo {author} {\bibfnamefont
  {W.}~\bibnamefont {Wegscheider}},\ and\ \bibinfo {author} {\bibfnamefont
  {F.}~\bibnamefont {Nichele}},\ }\href@noop {}  (\bibinfo
  {year} {2023}),\ \Eprint {https://arxiv.org/abs/2306.00866} {arXiv:2306.00866
  } \BibitemShut {NoStop}%
\bibitem [{\citenamefont {Beenakker}\ and\ \citenamefont {van
  Houten}(1991)}]{PhysRevLett.66.3056}%
  \BibitemOpen
  \bibfield  {author} {\bibinfo {author} {\bibfnamefont {C.~W.~J.}\
  \bibnamefont {Beenakker}}\ and\ \bibinfo {author} {\bibfnamefont
  {H.}~\bibnamefont {van Houten}},\ }\href
  {https://doi.org/10.1103/PhysRevLett.66.3056} {\bibfield  {journal} {\bibinfo
   {journal} {Phys. Rev. Lett.}\ }\textbf {\bibinfo {volume} {66}},\ \bibinfo
  {pages} {3056} (\bibinfo {year} {1991})}\BibitemShut {NoStop}%
\bibitem [{\citenamefont {Freyn}\ \emph {et~al.}(2011)\citenamefont {Freyn},
  \citenamefont {Dou{\c c}ot}, \citenamefont {Feinberg},\ and\ \citenamefont
  {M{\'e}lin}}]{freyn_production_2011}%
  \BibitemOpen
  \bibfield  {author} {\bibinfo {author} {\bibfnamefont {A.}~\bibnamefont
  {Freyn}}, \bibinfo {author} {\bibfnamefont {B.}~\bibnamefont {Dou{\c c}ot}},
  \bibinfo {author} {\bibfnamefont {D.}~\bibnamefont {Feinberg}},\ and\
  \bibinfo {author} {\bibfnamefont {R.}~\bibnamefont {M{\'e}lin}},\ }\href
  {https://doi.org/10.1103/PhysRevLett.106.257005} {\bibfield  {journal}
  {\bibinfo  {journal} {Physical Review Letters}\ }\textbf {\bibinfo {volume}
  {106}},\ \bibinfo {pages} {257005} (\bibinfo {year} {2011})} \BibitemShut {NoStop}%
\bibitem [{\citenamefont {Feinberg}\ \emph {et~al.}(2015)\citenamefont
  {Feinberg}, \citenamefont {Jonckheere}, \citenamefont {Rech}, \citenamefont
  {Martin}, \citenamefont {Dou{\c c}ot},\ and\ \citenamefont
  {M{\'e}lin}}]{feinberg_quartets_2015}%
  \BibitemOpen
  \bibfield  {author} {\bibinfo {author} {\bibfnamefont {D.}~\bibnamefont
  {Feinberg}}, \bibinfo {author} {\bibfnamefont {T.}~\bibnamefont
  {Jonckheere}}, \bibinfo {author} {\bibfnamefont {J.}~\bibnamefont {Rech}},
  \bibinfo {author} {\bibfnamefont {T.}~\bibnamefont {Martin}}, \bibinfo
  {author} {\bibfnamefont {B.}~\bibnamefont {Dou{\c c}ot}},\ and\ \bibinfo
  {author} {\bibfnamefont {R.}~\bibnamefont {M{\'e}lin}},\ }\href
  {https://doi.org/10.1140/epjb/e2015-50849-3} {\bibfield  {journal} {\bibinfo
  {journal} {The European Physical Journal B}\ }\textbf {\bibinfo {volume}
  {88}},\ \bibinfo {pages} {99} (\bibinfo {year} {2015})}\BibitemShut {NoStop}%
\bibitem [{\citenamefont {Sliwa}\ \emph {et~al.}(2015)\citenamefont {Sliwa},
  \citenamefont {Hatridge}, \citenamefont {Narla}, \citenamefont {Shankar},
  \citenamefont {Frunzio}, \citenamefont {Schoelkopf},\ and\ \citenamefont
  {Devoret}}]{sliwa_reconfigurable_2015}%
  \BibitemOpen
  \bibfield  {author} {\bibinfo {author} {\bibfnamefont {K.~M.}\ \bibnamefont
  {Sliwa}}, \bibinfo {author} {\bibfnamefont {M.}~\bibnamefont {Hatridge}},
  \bibinfo {author} {\bibfnamefont {A.}~\bibnamefont {Narla}}, \bibinfo
  {author} {\bibfnamefont {S.}~\bibnamefont {Shankar}}, \bibinfo {author}
  {\bibfnamefont {L.}~\bibnamefont {Frunzio}}, \bibinfo {author} {\bibfnamefont
  {R.~J.}\ \bibnamefont {Schoelkopf}},\ and\ \bibinfo {author} {\bibfnamefont
  {M.~H.}\ \bibnamefont {Devoret}},\ }\href
  {https://doi.org/10.1103/PhysRevX.5.041020} {\bibfield  {journal} {\bibinfo
  {journal} {Physical Review X}\ }\textbf {\bibinfo {volume} {5}},\ \bibinfo
  {pages} {041020} (\bibinfo {year} {2015})}\BibitemShut {NoStop}%
\bibitem [{\citenamefont {Frattini}\ \emph {et~al.}(2017)\citenamefont
  {Frattini}, \citenamefont {Vool}, \citenamefont {Shankar}, \citenamefont
  {Narla}, \citenamefont {Sliwa},\ and\ \citenamefont
  {Devoret}}]{frattini_3-wave_2017}%
  \BibitemOpen
  \bibfield  {author} {\bibinfo {author} {\bibfnamefont {N.~E.}\ \bibnamefont
  {Frattini}}, \bibinfo {author} {\bibfnamefont {U.}~\bibnamefont {Vool}},
  \bibinfo {author} {\bibfnamefont {S.}~\bibnamefont {Shankar}}, \bibinfo
  {author} {\bibfnamefont {A.}~\bibnamefont {Narla}}, \bibinfo {author}
  {\bibfnamefont {K.~M.}\ \bibnamefont {Sliwa}},\ and\ \bibinfo {author}
  {\bibfnamefont {M.~H.}\ \bibnamefont {Devoret}},\ }\href
  {https://doi.org/10.1063/1.4984142} {\bibfield  {journal} {\bibinfo
  {journal} {Applied Physics Letters}\ }\textbf {\bibinfo {volume} {110}},\
  \bibinfo {pages} {222603} (\bibinfo {year} {2017})}\BibitemShut {NoStop}%
\bibitem [{\citenamefont {Frattini}\ \emph {et~al.}(2018)\citenamefont
  {Frattini}, \citenamefont {Sivak}, \citenamefont {Lingenfelter},
  \citenamefont {Shankar},\ and\ \citenamefont
  {Devoret}}]{frattini_optimizing_2018}%
  \BibitemOpen
  \bibfield  {author} {\bibinfo {author} {\bibfnamefont {N.~E.}\ \bibnamefont
  {Frattini}}, \bibinfo {author} {\bibfnamefont {V.~V.}\ \bibnamefont {Sivak}},
  \bibinfo {author} {\bibfnamefont {A.}~\bibnamefont {Lingenfelter}}, \bibinfo
  {author} {\bibfnamefont {S.}~\bibnamefont {Shankar}},\ and\ \bibinfo {author}
  {\bibfnamefont {M.~H.}\ \bibnamefont {Devoret}},\ }\href
  {https://doi.org/10.1103/PhysRevApplied.10.054020} {\bibfield  {journal}
  {\bibinfo  {journal} {Physical Review Applied}\ }\textbf {\bibinfo {volume}
  {10}},\ \bibinfo {pages} {054020} (\bibinfo {year} {2018})}\BibitemShut
  {NoStop}%
\bibitem [{\citenamefont {Matsuo}\ \emph {et~al.}(2023)\citenamefont {Matsuo},
  \citenamefont {Imoto}, \citenamefont {Yokoyama}, \citenamefont {Sato},
  \citenamefont {Lindemann}, \citenamefont {Gronin}, \citenamefont {Gardner},
  \citenamefont {Manfra},\ and\ \citenamefont {Tarucha}}]{matsuo2023josephson}%
  \BibitemOpen
  \bibfield  {author} {\bibinfo {author} {\bibfnamefont {S.}~\bibnamefont
  {Matsuo}}, \bibinfo {author} {\bibfnamefont {T.}~\bibnamefont {Imoto}},
  \bibinfo {author} {\bibfnamefont {T.}~\bibnamefont {Yokoyama}}, \bibinfo
  {author} {\bibfnamefont {Y.}~\bibnamefont {Sato}}, \bibinfo {author}
  {\bibfnamefont {T.}~\bibnamefont {Lindemann}}, \bibinfo {author}
  {\bibfnamefont {S.}~\bibnamefont {Gronin}}, \bibinfo {author} {\bibfnamefont
  {G.~C.}\ \bibnamefont {Gardner}}, \bibinfo {author} {\bibfnamefont {M.~J.}\
  \bibnamefont {Manfra}},\ and\ \bibinfo {author} {\bibfnamefont
  {S.}~\bibnamefont {Tarucha}},\ }\href@noop {}  (\bibinfo {year}
  {2023}),\ \Eprint {https://arxiv.org/abs/2305.07923} {arXiv:2305.07923} \BibitemShut {NoStop}%
\end{thebibliography}
\end{document}